\documentclass{article}

\usepackage{arxiv}

\usepackage[utf8]{inputenc} 
\usepackage[T1]{fontenc}    
\usepackage{hyperref}       
\usepackage{url}            
\usepackage{booktabs}       
\usepackage{amsfonts}       
\usepackage{nicefrac}       
\usepackage{microtype}      
\usepackage{lipsum}
\usepackage{graphicx}
\usepackage{multirow}
\usepackage{epstopdf}
\usepackage{subfigure}

\title{Experimental and numerical study for the representation of standing waves in a Kundt's tube}

\author{
  Alex Estupiñán\thanks{djlexes@gmail.com, personal email address}\\
  Facultad de Ciencias, Escuela de Física\\
  Universidad Industrial de Santander (UIS)\\
  Colombia, Santander, Bucaramanga \\
  \texttt{alex.estupinan@saber.uis.edu.co} \\
   \And
      Raul Ortiz \\
      Departamento De Matemáticas Y Ciencias Naturales\\
     Universidad Autónoma de Bucaramanga (UNAB)\\
      Colombia, Santander, Bucaramanga\\
      \texttt{rortiz579@unab.edu.co} \\
   \And
   Ángel Duarte, Jesús Sanchez, Jefferson Arciniegas \\
   Universidad Autónoma de Bucaramanga (UNAB)\\
   Colombia, Santander, Bucaramanga\\
}

\begin{document}
\maketitle

\begin{abstract}

The study of the behavior of a standing wave inside a resonant cavity, such as a closed tube, has always been of great revelation in the manufacture of wind musical instruments, in addition to research carried out on the transfer and absorption of heat, energy and pressure for the acoustic characterization of materials by injecting sound waves at one end of the Kundt's tube.

In this work, we did the implementation of two experiments for measure the speed of sound, using a Kundt's tube where also we obtained the bellies and nodes of the sound wave and we verificated with a computational code that was wrote for us in Python, in which we made simulation of the temporal evolution was achieved of the wave that vibrated inside the tube. 

Finally, we show the wave profile for the first four harmonics with computational simulations and we presented the experimental results by two different methods, these were compared with the obtained using the equations modeled by the undulatory theory of physics.

\end{abstract}


\section{Introduction}

Undulatory phenomena have been of great importance in the study of physics, for teachers, researchers and engineers in many fields such as optics and acoustics, in which these phenomena have several relevant applications. Using the fundamental concepts, belonging to the branch of physics waves and particles, we present an experimental procedure, with which the definition of a standing wave can be explained, as the superposition of two waves propagating in opposite directions, besides being able to show the position of the nodes and bellies using small spheres of polystyrene inside the tube (See the Figure \ref{fig_3}). 

With the purpose of model the equations that govern the movement of the wave inside the tube, we realiced a computational simulation, with wich we show the change in the behavior of the wave for different frequency values, in order to better observe the vibratory movement of the standing wave present in the tube, here it was possible to identify and verify the corresponding independence of the number of nodes and antinodes as a function of the oscillation fundamental frequency and the harmonics of the acustic wave.

The remaining part of this paper is organized as follows: Section \ref{model}, describes the theoretical background model corresponding to the calculates that we needing for modeling the phenomenom to study. In Section \ref{experimental}, shown the experimental procedure for the data-taking. In Section \ref{results},  presents the results of the numerical analysis  for the four first harmonic, that is developed with a series of simulations, after of this we showed the experimental results of calculation sound velocity in the Kundt’s tube, using two different methods.

\section{Theoretical Background}
\label{model}

\begin{figure}[h!]
	\centering
	\includegraphics[width=0.95\textwidth]{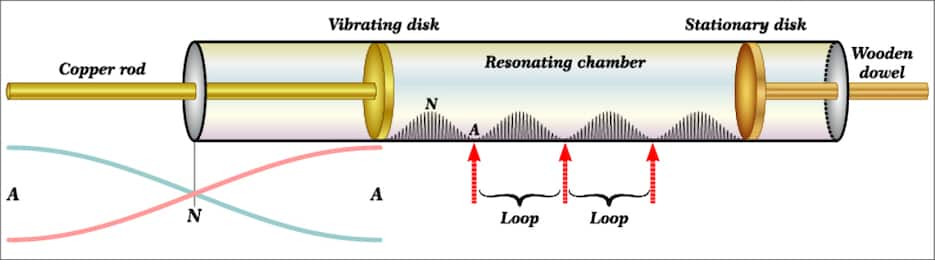}
	\caption{Physical scheme of system Kundt's tube, where $A$ indicates the position of the antinodes and $N$ the position of the nodes.}
	\label{fig_1}
\end{figure}

The physical system to be studied is shown in the Figure \ref{fig_1}.

The theoretical basis for understanding the operation and being able to interpret the experimental results with the Kundt's tube focuses on the study of standing waves and their discretization in normal modes \cite{Benson}, stationary waves are a particular case of the phenomenon of wave interference, since it deals with the superposition of waves with equal amplitudes and wavelengths, which travel in the same direction but in the opposite directions \cite{Yunes}.

In the standing waves in tubes, each gas molecule oscillates around its equilibrium position when the tube is excited at a certain frequency.  In the case of the Kundt's tube, the standing waves are inside a tube that usually has one end closed. When the tube is closed at both ends it is called a closed tube. 
In addition, it is of great importance to emphasize that the amplitude of the resulting stationary wave generated, despreciando the effects of absorption and damping, must be twice the amplitude of the waves that initially overlap to form it. The equation that models the oscillatory behavior of a standing wave is expressed in the Equation (\ref{eq_1})

\begin{equation}
 y(x,t) = 2 A sin(kx)\cdot cos(\omega t),
 \label{eq_1}
\end{equation} 

where $k$ and $\omega$, are the number of wave and the angular frequency respectively, their mathematical expression is given by: 

\begin{equation}
	k=\frac{2\pi}{\lambda} \quad and \quad \omega = 2\pi f 
\end{equation}

where $\lambda$ is the wavelength, in order to find the position of the nodes and bellies inside the tube, we must take into account Equation (\ref{eq_1}), where to find the nodal points of zero oscillation, we do:

\begin{equation}
	sin \left( \frac{2\pi}{\lambda} x \right) = 0.
\end{equation}

The condition for the sinusoidal function to be canceled is fulfilled only for $n$ integer values of $\pi$

\begin{equation}
	\frac{2\pi}{\lambda} x = n\pi,
\end{equation}

from the above equation it follows that the standing wave presents a node for the position $x$:

\begin{equation}
	x = \frac{n \lambda}{2}
	\label{nodo}
\end{equation}

It is also very important to find the position of the bellies of the wave, which will be the values of maximum amplitude of the stationary wave, algebraically this is expressed as:

\begin{equation}
	sin \left( \frac{2\pi}{\lambda} x \right) = \pm 1,
\end{equation}

The condition for the function $sin(kx)$ to be maximum is given for odd integer values $ 2n + 1 $, in the following way:

\begin{equation}
\frac{2\pi}{\lambda} x	= \frac{(2n+1) \pi}{2}.
\end{equation}

From the above equation, we can deduce that the position for each belly is:

\begin{equation}
	x = (2n+1)\frac{\lambda}{4}
	\label{vientre}	
\end{equation}

From equations (\ref{nodo}) and (\ref{vientre}) it can be seen that the separation distance between two nodes and two consecutive bellies is equal to half a wavelength $\lambda/2$.

Continuing with the boundary conditions of a Kundt's tube, for which at the two ends of the tube, a node is presented,

\begin{equation}
	y(0,t) = 0 \quad and \quad y(L,t) = 0.
\end{equation}

Where L is the length of the tube ,applying these boundary conditions to equation (\ref{eq_1}), we obtain the $n$ normal modes of system oscillation, which are:

\begin{equation}
	 2 A sin(kL)\cdot cos(\omega t) = 0  
\end{equation}

continuing in the operation for the number wave $k$, we obtained 

\begin{equation}
	sin(kL) = 0, \quad k_n = \frac{n \pi}{L},  
\end{equation}

where $n$ are integers values, the above equation as a function of the wavelength, can be expressed as follows,

\begin{equation}
\lambda_n = \frac{2\pi}{k_n} = \frac{2L}{n}.	
\end{equation}

If we use the ratio of the speed of propagation of the wave, depending on the wavelength and frequency, we can obtain the frequency for the different harmonics $n$, in the following way:

\begin{equation}
	v = \lambda_n \cdot f, \quad f_n = \frac{nv}{2L},
	\label{velocidad}
\end{equation}

with this last expression, we can calculate the speed of sound inside the resonant chamber,

\begin{equation}
	v_n = \frac{2L}{n} \cdot f	
\label{velocidad_2}
\end{equation}

\section{Experimental study}
\label{experimental}

\begin{figure}[h!]
	\centering
	\subfigure[]
	{
		\includegraphics[width=0.47\linewidth]{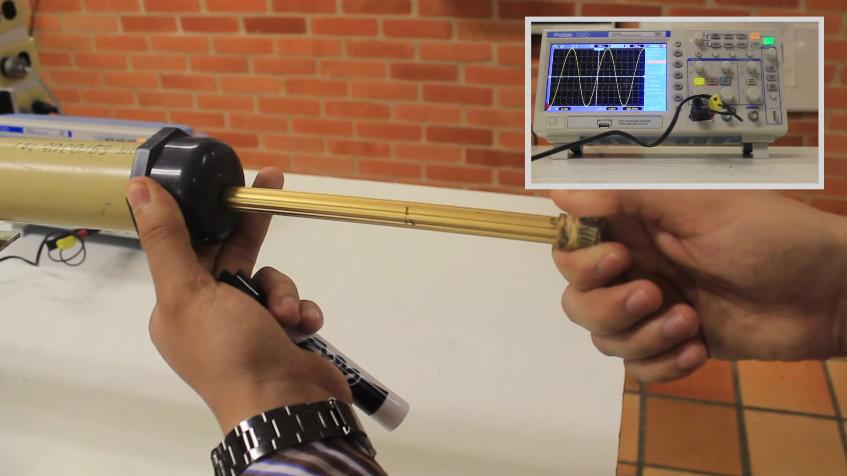}
			\label{fig_2}
	}
	\subfigure[]
	{
		\includegraphics[width=0.49\linewidth]{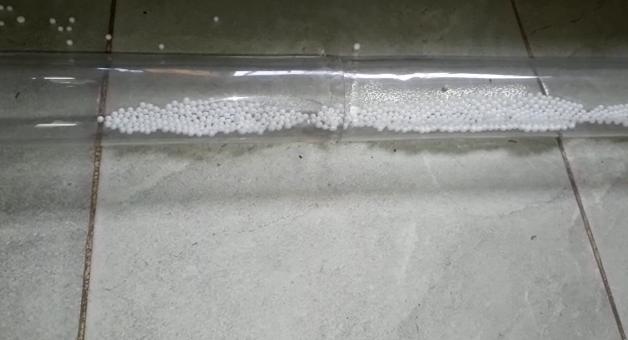}
		\label{fig_3}
	}
		\caption{(a) Experimental setup of a Kundt's tube, using a PVC tube and an oscilloscope. (b) Experimental setup of a Kundt's tube, using an acrylic tube and small spheres of polystyrene inside the chamber.}
\end{figure}

To perform the experimental study of standing waves in the Kundt's tubes, the assemblies shown in figures \ref{fig_2} and \ref{fig_3} were made, where in the experiment of Figure \ref{fig_2}, it was necessary to use a digital oscilloscope to detect the position of the antinodal points and in Figure \ref{fig_3} this detection was used observing the organization of the small polystyrene spheres \cite{alba}.

For both experiments, we used an signal generator and mesuremented the separation distance between two consecutive bellies (antinodes) $L_x$, also we registered the wavelength value $\lambda$ and period $T$, were recorded for six different frequencies values, these data are recorded in the tables \ref{table_1} and \ref{table_2}.

\begin{table}[h!]
	\centering
	\begin{tabular}{|c|c|c|c|cllll}
		\cline{1-4}
		\multicolumn{1}{|l|}{\textbf{Frequency {[}hz{]}}} & \multicolumn{1}{l|}{\textbf{$L_x$ {[}m{]}}} & \multicolumn{1}{l|}{\textbf{$\lambda$ {[}m{]}}} & \multicolumn{1}{l|}{\textbf{T {[}ms{]}}} & \multicolumn{1}{l}{\textbf{}} &  &  &  &  \\ \cline{1-4}
		600                                               & 0.288                                       & 0.576                                           & 1.666                                    &                               &  &  &  &  \\ \cline{1-4}
		800                                               & 0.215                                       & 0.430                                           & 1.250                                    &                               &  &  &  &  \\ \cline{1-4}
		1000                                              & 0.173                                       & 0.346                                           & 1                                        &                               &  &  &  &  \\ \cline{1-4}
		1200                                              & 0.142                                       & 0.284                                           & 0.833                                    &                               &  &  &  &  \\ \cline{1-4}
		1400                                              & 0.127                                       & 0.254                                           & 0.714                                    &                               &  &  &  &  \\ \cline{1-4}
		1600                                              & 0.112                                       & 0.224                                           & 0.625                                    &                               &  &  &  &  \\ \cline{1-4}
	\end{tabular}
	\vspace{0.2cm}
	\caption{Experimental data, using the assembly of the Figure \ref{fig_2}.}
	\label{table_1}
\end{table}

\begin{table}[h!]
	\centering
	\begin{tabular}{|c|c|c|c|cllll}
		\cline{1-4}
		\multicolumn{1}{|l|}{\textbf{Frequency {[}hz{]}}} & \multicolumn{1}{l|}{\textbf{$L_x$ {[}m{]}}} & \multicolumn{1}{l|}{\textbf{$\lambda$ {[}m{]}}} & \multicolumn{1}{l|}{\textbf{T {[}ms{]}}} & \multicolumn{1}{l}{\textbf{}} &  &  &  &  \\ \cline{1-4}
		600                                               & 0.275                                       & 0.550                                           & 1.666                                    &                               &  &  &  &  \\ \cline{1-4}
		800                                               & 0.221                                       & 0.442                                           & 1.250                                    &                               &  &  &  &  \\ \cline{1-4}
		1000                                              & 0.178                                       & 0.356                                           & 1                                        &                               &  &  &  &  \\ \cline{1-4}
		1200                                              & 0.143                                       & 0.286                                           & 0.833                                    &                               &  &  &  &  \\ \cline{1-4}
		1400                                              & 0.129                                       & 0.258                                           & 0.714                                    &                               &  &  &  &  \\ \cline{1-4}
		1600                                              & 0.114                                       & 0.248                                           & 0.625                                    &                               &  &  &  &  \\ \cline{1-4}
	\end{tabular}
	\vspace{0.2cm}
	\caption{Experimental data, using the assembly of the Figure \ref{fig_3}.}
	\label{table_2}
\end{table}


\newpage

\section{Results}
\label{results}

\subsection{Numerical Analysis}
\label{numerical}

One of the relevant features of this work consists in the elaboration of a computer code written in Python, and executed using the Jupyter Notebook tool \cite{jupyter}, with which it was possible to observe the different harmonics for the case of a tube of 1 meter in length.

We did the simulation of the standing wave in a kundt's tube, for the first five harmonics using the equations (\ref{velocidad}) and (\ref{velocidad_2}) as shown in Table \ref{table_4}, where we take the speed of sound equal to $243$ m/s \cite{costa}.

\begin{table}[h!]
	\centering
	\begin{tabular}{ccccllll}
		\cline{1-4}
		\multicolumn{1}{|c|}{\textbf{Length of tube}}     & \multicolumn{3}{c|}{\textbf{1 m}}                                                                                                   &  &  &  &  \\ \cline{1-4}
		\multicolumn{1}{|c|}{\textbf{Frequancy {[}hz{]}}} & \multicolumn{1}{c|}{\textbf{n}} & \multicolumn{1}{c|}{\textbf{$\lambda$ {[}m{]}}} & \multicolumn{1}{c|}{\textbf{Period T {[}ms{]}}} &  &  &  &  \\ \cline{1-4}
		\multicolumn{1}{|c|}{171}                         & \multicolumn{1}{c|}{1}          & \multicolumn{1}{c|}{2}                          & \multicolumn{1}{c|}{5.847}                      &  &  &  &  \\ \cline{1-4}
		\multicolumn{1}{|c|}{343}                         & \multicolumn{1}{c|}{2}          & \multicolumn{1}{c|}{1}                          & \multicolumn{1}{c|}{2.915}                      &  &  &  &  \\ \cline{1-4}
		\multicolumn{1}{|c|}{514}                         & \multicolumn{1}{c|}{3}          & \multicolumn{1}{c|}{0.66}                       & \multicolumn{1}{c|}{1.945}                      &  &  &  &  \\ \cline{1-4}
		\multicolumn{1}{|c|}{686}                         & \multicolumn{1}{c|}{4}          & \multicolumn{1}{c|}{0.50}                       & \multicolumn{1}{c|}{1.457}                      &  &  &  &  \\ \cline{1-4}
		\multicolumn{1}{|c|}{857}                         & \multicolumn{1}{c|}{5}          & \multicolumn{1}{c|}{0.40}                       & \multicolumn{1}{c|}{1.167}                      &  &  &  &  \\ \cline{1-4} 
	\end{tabular}
	\vspace{0.2cm}
	\caption{Number of harmonics present in the tube as a function of frequency and period.}
	\label{table_4}
\end{table}

\begin{figure}[h]
	\begin{center}
		\begin{tabular}{cc}
			
			\includegraphics[scale=16.29]{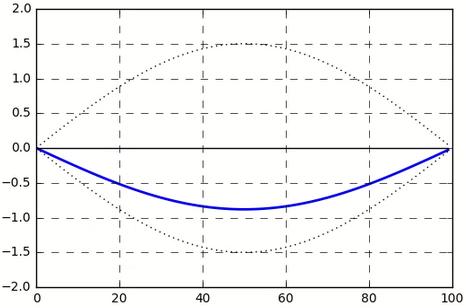} &
			\includegraphics[scale=16.29]{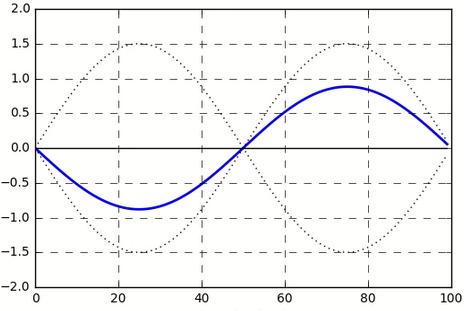}
			\\ (a) & (b)
			\\ 
			\includegraphics[scale=16.29]{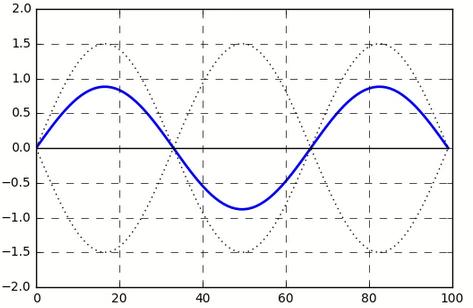} &
			\includegraphics[scale=18.29]{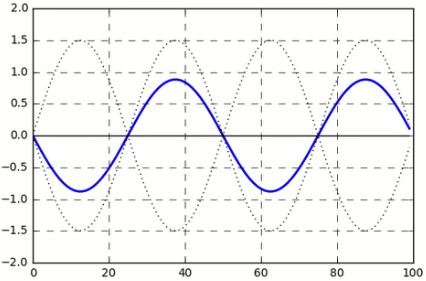}
			\\ (c) & (d)
			\\ 
		\end{tabular}
		\caption{\label{fig_total}
			Behavior of the standing wave for the $n$ first four harmonics, on the $X$ axis is the length of the tube and on the $Y$ axis the amplitude of the wave: (a) n = 1. (b) n = 2. (c) n = 3. (d) n = 4.}
	\end{center}
\end{figure}

In the figure \ref{fig_total}, we can observe the behavior of the standing wave inside the Kundt's tube for the first four harmonics.

\subsection{Experimental Analysis}
\label{experimental_results}

In order to measure the speed of sound in the air \cite{Nursulistiyo}, using the two experimental assemblies shown in the section \ref{experimental}, we look for an analytical model that can be adjusted to the experimental data presented in tables \ref{table_1} and \ref{table_2}.

\begin{figure}[h!]
	\centering
	\includegraphics[width=0.58\textwidth]{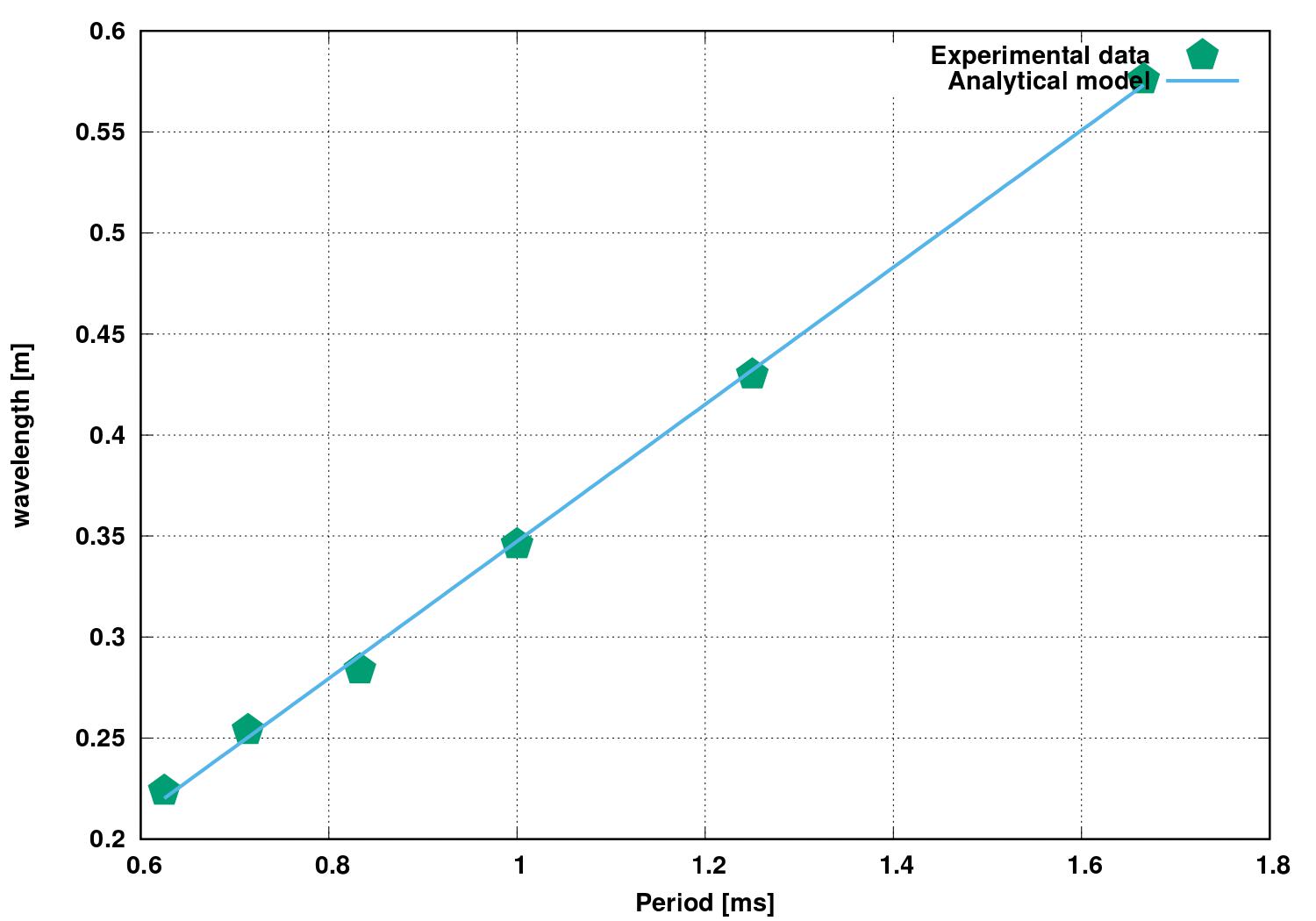}
	\caption{Linear fitting for the experiment in which we use the PVC tube.}
	\label{fig_4}
\end{figure}

We used the values in Table \ref{table_1} and performing a linear fit of the data with the Equation (\ref{velocidad_esperimental}), we obtained the relationship shown in Figure \ref{fig_4}.

\begin{equation}
	f = 1/T, \quad v = \lambda \cdot f \quad \Rightarrow \quad \lambda = v \cdot T
	\label{velocidad_esperimental}
\end{equation}

\begin{figure}[h!]
	\centering
	\includegraphics[width=0.58\textwidth]{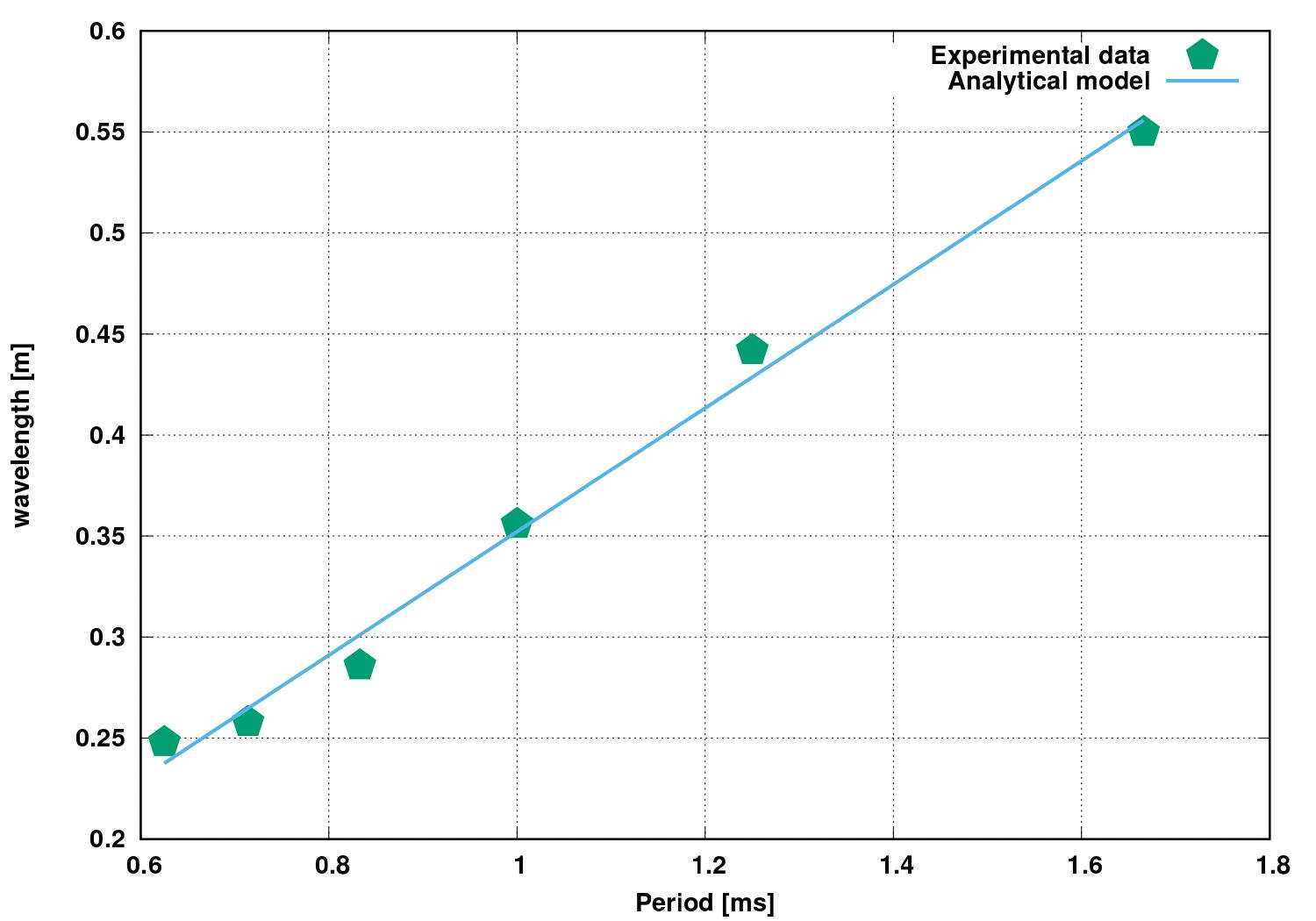}
	\caption{Linear fitting for the experiment in which we use the acrylic tube.}
	\label{fig_5}
\end{figure}

On the other hand, the linear fit for the data shown in Table \ref{table_2} is shown in Figure \ref{fig_5},

using the linear model shown in Equation (\ref{velocidad_esperimental}), Table \ref{table_3} was constructed, in which we can compare the experimental error obtained for both types of experimental assemblies, where the values obtained by these two experimental adjustments were compared with the speed of sound at 20 degrees Celsius ($v  \approx 343.2$  $m/s$)

\newpage

\begin{table}[h!]
	\centering
	\begin{tabular}{|c|c|c|c|}
		\hline
		\textbf{Experimental setup} & \textbf{Speed of sound {[}m/s{]}} & \textbf{Absolute error {[}m/s{]}} & \textbf{\% Relative error} \\ \hline
		No.1 (tube of PVC)          & 339.099                           & 4.101                             & 1.194                      \\ \hline
		No.2 (tube of acrylic)      & 305.749                           & 37.451                            & 10.912                     \\ \hline
	\end{tabular}
	\vspace{0.2cm}
		\caption{Experimental errors obtained for the two experimental configurations implemented.}
		\label{table_3}
\end{table} 

From the results shown in Table \ref{table_3}, we could notice that a lower percentage of error is obtained for the case of the first experimental assembly, in which we use the digital oscilloscope, unlike the second experimental setup, in which we observe the spatial distribution of the small spheres of polystyrene, for measured the distance between two consecutive bellies.

\section{Conclusions}

In this article, we were able to show two experimental methods, in which we were able to measure the speed of sound, with an experimental percentage relative error below 11\%, also we present the results obtained from the computational simulation; for the first four harmonics and with this, we can understand better the wave phenomenon that occurs inside a Kundt tube.

\section*{Acknowledgments}

The authors would like to thank the Universidad Autónoma de Bucaramanga (UNAB), for lend us the installations and materials for realizing these two experiments presented in this paper.


\begin{thebibliography}{unsrt}

\bibitem{Benson} Benson, H. (2000). Física universitaria, Vol. 1. México: Cecsa.
\bibitem{Yunes}Yunes, C. A., \& Boles, M. A. (2012). Termodinámica. editorial MacGrawHill
\bibitem{alba} Alba, J., Marant, V., Aguilera, J. L., \& Ramis, J. (2006). Criterios de selección de materiales acústicos absorbentes con técnicas basadas en tubo de Kundt. TecniAcústica, Gandía.
\bibitem{jupyter} Kluyver, T., Ragan-Kelley, B., Pérez, F., Granger, B. E., Bussonnier, M., Frederic, J., ... \& Ivanov, P. (2016, May). Jupyter Notebooks-a publishing format for reproducible computational workflows. In ELPUB (pp. 87-90).
\bibitem{costa} da Costa Saab, S., Cássaro, F. A. M., \& Brinatti, A. M. (2005). Laboratório caseiro: tubo de ensaio adaptado como tubo de kundt para medir a velocidade do som no ar. Caderno Brasileiro de Ensino de Física, 22(1), 112-120.
\bibitem{Nursulistiyo} Nursulistiyo, E. (2018, March). Design and development of multipurpose Kundt’s tube as physics learning media. In Journal of Physics: Conference Series (Vol. 983, No. 1, p. 012011). IOP Publishing.


\end{thebibliography}
\end{document}